\documentclass[12pt,preprint]{aastex}
\usepackage{textcomp}
\usepackage{amsmath}
\usepackage{graphicx}
\usepackage[T1]{fontenc}

\def\beq{\begin{eqnarray}}
\def\eeq{\end{eqnarray}}

\begin{document}

\title{Can black-hole neutrino-cooled disks power short gamma-ray bursts?}

\author{Tong Liu\altaffilmark{1,2,3,4,*}, Yi-Qing Lin\altaffilmark{5}, Shu-Jin Hou\altaffilmark{6,2}, Wei-Min Gu\altaffilmark{1,4}}

\altaffiltext{1}{Department of Astronomy and Institute of Theoretical Physics and Astrophysics, Xiamen University, Xiamen, Fujian 361005, China}
\altaffiltext{2}{Key Laboratory for the Structure and Evolution of Celestial Objects, Chinese Academy of Sciences, Kunming, Yunnan 650011, China}
\altaffiltext{3}{State Key Laboratory of Theoretical Physics, Institute of Theoretical Physics, Chinese Academy of Sciences, Beijing 100190, China}
\altaffiltext{4}{SHAO-XMU Joint Center for Astrophysics, Xiamen University, Xiamen, Fujian 361005, China}
\altaffiltext{5}{School of Opto-electronic and Communication Engineering, Xiamen University of Technology, Xiamen, Fujian 361024, China}
\altaffiltext{6}{College of Physics and Electronic Engineering, Nanyang Normal University, Nanyang, Henan 473061, China}
\email{*tongliu@xmu.edu.cn}

\begin{abstract}
Stellar-mass black holes (BHs) surrounded by neutrino-dominated accretion flows (NDAFs) are the plausible candidates to power gamma-ray bursts (GRBs) via neutrinos emission and their annihilation. The progenitors of short-duration GRBs (SGRBs) are generally considered to be compact binaries mergers. According to the simulation results, the disk mass of the NDAF has been limited after merger events. We can estimate such disk mass by using the current SGRB observational data and fireball model. The results show that the disk mass of a certain SGRB mainly depends on its output energy, jet opening angle, and central BH characteristics. Even for the extreme BH parameters, some SGRBs require massive disks, which approach or exceed the limits in simulations. We suggest that there may exist alternative magnetohydrodynamic processes or some mechanisms increasing the neutrino emission to produce SGRBs with the reasonable BH parameters and disk mass.
\end{abstract}

\keywords {accretion, accretion disks - black hole physics - gamma-ray burst: general - neutrinos}

\section{Introduction}

Gamma-ray bursts (GRBs) are the most powerful electromagnetic events in the universe, which are sorted into two categories, i.e., short- and long-duration GRBs \citep[SGRBs and LGRBs, see][]{Kouveliotou1993} or type I and II GRBs \citep{Zhang2006,Zhang2007a}. Their progenitors are considered to be mergers of two compact objects, i.e., two neutron stars (NSs) or a black hole (BH) and a NS \citep[for reviews, see, e.g.,][]{Nakar2007,Berger2014}, and collapses of massive stars (e.g., \citealt{Woosley2006} for reviews), respectively. For the interpretations on the gamma-ray and afterglow emission of GRBs, the fireball shock model \citep[for reviews, see, e.g.,][]{Meszaros2002,Zhang2004} has been widely accepted. The popular models on the central engines of GRBs are either a rotating stellar BH surrounded by a hyperaccretion disk \citep[e.g.,][]{Paczynski1991,Narayan1992,MacFadyen1999} or a quickly rotating magnetar \citep[or protomagnetar, e.g.,][]{Usov1992,Metzger2011,Lv2015}.

Two mechanisms have been proposed to power GRBs if a hyperaccretion disk exists in the center of GRBs, i.e., neutrino emission and annihilation, and magnetohydrodynamic processes, such as Blandford-Znajek (BZ) mechanism \citep{Blandford1977} and episodic magnetic reconnection \citep{Yuan2012}. For the former mechanism, neutrino annihilation can produce a relativistic electron-positron outflow, which is considered as the progenitor of the fireball to power a GRB. The most probable model to launch a large number of neutrinos is a geometrically and optically thick neutrino-cooled hyperaccretion disks, named as neutrino-dominated accretion flow (NDAF), whose typical characteristics are extremely high accretion rate and neutrino-cooling process. In the inner region of the NDAF, the main components are the electrons, free neutrons and protons, the density and temperature are very high ($\rho \sim 10^{10}-10^{13}~\rm g~cm^{-3}$ and $T \sim ~10^{10}-10^{11}~\rm K$), and the photons are tightly trapped in such disk, thus the energy loss is mainly through neutrino and antineutrino radiation \citep[see, e.g.,][]{Popham1999,DiMatteo2002,Kohri2002,Rosswog2002,Kohri2005,Lee2005,Gu2006,Chen2007,Liu2007,Kawanaka2007,Zalamea2011,Janiuk2013,Kawanaka2013,Xue2013}.

Two factors should be considered in the calculation of the neutrino luminosity and annihilation luminosity, which are the structure and components of the NDAF and the description of the relativistic neutrino propagation. \citet{Xue2013} investigated the global solutions of the radial structure and components of the NDAF in the Kerr space-time of the BH with detailed neutrino physics and nucleosynthesis processes. The results show that the gas pressure and the neutrino cooling are always dominant in the inner region for the high mass accretion rate, and the major components of the inner, middle, and outer regions are the free nucleons, $\rm ^4He$, and $\rm ^{56}Fe$, respectively. Importantly, they noticed that the radiative neutrinos mainly come from the inner region of the disk, and the neutrino emission rate less depends on the description of the microphysics, as well as other studies of NDAF model \citep[e.g.,][]{Popham1999,DiMatteo2002,Liu2007,Kawanaka2013}. Even for the discussions on the vertical structure of the NDAF, the similar solutions are presented \citep[e.g.,][]{Liu2008,Liu2010,Liu2012a,Liu2013,Liu2014}. Thus the main problem is how to precisely calculate the neutrino annihilation processes. \citet{Birkl2007}, \citet{Kovacs2011a}, and \citet{Kovacs2011b} analyzed the influence of general relativistic effects on the neutrino annihilation efficiency, which has a prominent increase compared with the Newtonian calculations. Based on the geodesic-tracing method, \citet{Zalamea2011} also studied annihilations via tracing the neutrino track.

For SGRBs, \citet{Eichler1989} proposed that the mergers of two NSs might be the candidates. \citet{Ruffert1998} simulated three-dimensional Newtonian hydrodynamical solutions of the merger events of two NSs with mass $\sim 1.6~M_\odot$. There might survive a disk $\sim 0.1-0.2~M_\odot$ surrounding a BH $\sim 2.5~M_\odot$. Furthermore, \citet{Paczynski1991} and \citet{Narayan1992} presented that the merger of a NS and a stellar-mass BH can also produce a SGRB. In simulations, the fragments of the NS can form a more massive disk, $\sim 0.5 M_\odot$ \citep[e.g.,][]{Kluzniak1998,Lee1999,Popham1999,Liu2012b}. In the past several years, the massive NSs, i.e., $\sim 2~M_\odot$ have been discovered in binaries, which accompany the white dwarfs \citep{Demorest2010,Antoniadis2013}. Yet we cannot neglect the possibilities that the massive NSs exist in the BH-NS or two NSs binaries. However, the mass of the disk is still much smaller than $1~M_\odot$ with the logical conjecture. So it begs a question: can annihilations of neutrinos from NDAFs owning such disk masses power all the observed SGRBs? \citet{Fan2011} investigated the disk mass in the center of SGRBs with the fixed values of the BH mass and spin. They found that nearly half SGRBs are suitable for the results of the above simulations. We further consider that the annihilation description, intact samples with prompt emission and afterglow properties of SGRBs, and reasonable ranges of the BH parameters should be fully included to answer the above question.

In Section 2, we describe the physical processes from the neutrino annihilation to observational gamma-ray photons. By using the current SGRBs data, the disk masses for the definite ranges of the BH parameters are shown in Section 3. Conclusions and discussion are in Section 4.

\section{Model}

The neutrino annihilation luminosity $L_{\nu\bar{\nu}}$ is a function of the BH mass $M_{\rm BH} $, dimensionless spin parameter $a_*$ ($a_* \equiv c J/GM_{_{\rm BH}}^{2}$, $J$ is the angular momentum of the BH), dimensionless viscosity parameter $\alpha$, and mass accretion rate $\dot{M}$ \citep[see, e.g.,][]{Popham1999,Rosswog2003,Gu2006,Liu2007,Zalamea2011,Kawanaka2013,Xue2013,Leng2014}.

The analytical formula of $L_{\nu\bar{\nu}}$ is shown in many previous works \citep[e.g.,][]{Fryer1999,Zalamea2011,Xue2013}. Here we adopt the neutrino annihilation luminosity $L_{\nu\bar{\nu}}$ given by \citet{Zalamea2011}, which is expressed as
\beq
L_{\nu\bar{\nu}}&\approx&5.7 \times 10^{52} ~x_{\rm ms}^{-4.8}~({M_{_{\rm BH}} / M_\odot})^{-3/2} \nonumber \\
&&\times \Bigg \{
\begin{array}{ll}
0 & \hbox{for $\dot{M} < \dot {M}_{\rm ign}$}\\
(\dot{M} / M_\odot~\rm s^{-1})^{9/4} & \hbox{for $\dot {M}_{\rm ign} < \dot{M} < \dot{M}_{\rm trap}$}\\
(\dot{M}_{\rm trap} / M_\odot~\rm s^{-1})^{9/4} & \hbox{for $\dot{M} > \dot{M}_{\rm trap}$}\\
\end{array}
\Bigg \} ~\rm erg~s^{-1},
\eeq
where $x_{\rm ms}=r_{\rm ms}/r_{\rm g}$, $r_{\rm ms}$ is radius of the last (marginally stable) orbit, $r_{\rm g}=2GM_{\rm BH}/c^2$ is the Schwarzschild radius, and $\dot {M}_{\rm ign}$ is the critical ignition accretion rate, $\dot {M}_{\rm trap}$ is the accretion rate if neutrino trapping events occur in the inner region of the NDAF \citep[e.g.,][]{DiMatteo2002,Kohri2005,Chen2007,Liu2012a,Xue2013}. Their numerical results depend on the viscosity parameter $\alpha$ and BH spin parameter $a_*$. Additionally, the value of viscosity parameter $\alpha$ has little effects on $L_{\nu\bar{\nu}}$ as long as $\dot {M}_{\rm ign} < \dot{M} < \dot{M}_{\rm trap}$ \citep{Zalamea2011}, so $\alpha = 0.1$ is adopted here. Furthermore, $x_{\rm ms}$ can be expressed as \citep[e.g.,][]{Bardeen1972,Kato2008,Hou2014}
\beq
x_{\rm ms}=\frac{1}{2} [3+Z_{2}- \sqrt{(3-Z_{1})(3+Z_{1}+2Z_{2})}],
\eeq
where
\beq
Z_{1}=1+(1-a_*^{2})^{1/3}[(1+a_*)^{1/3}+(1-a_*)^{1/3}],
\eeq
\beq
Z_{2}=\sqrt{3a_*^{2}+Z_{1}^{2}}.
\eeq
In comparison, \citet{Xue2013} also gave a similar analytical solution, i.e., $L_{\nu\bar{\nu}} \varpropto \dot{M}^{2.17}$, but the influence of the BH mass was not considered.

\citet{Popham1999} and \citet{Liu2007} investigated the spatial distribution of neutrino annihilation rate and found that nearly 60\% of the total annihilation luminosity is ejected from the region $r < 20~ r_{\rm g}$. In the studies on the vertical structure of NDAF model \citep[e.g.,][]{Liu2010,Liu2012a,Liu2013}, we found that the half-opening angle of the disk is very large, $\gtrsim 80^\circ$, for the typical accretion rate, $\sim 1~M_\odot~\rm s^{-1}$, thus the empty funnel along the rotation axis above the disk can naturally limit the opening angle of the neutrino annihilable ejection to produce the primary fireball.

The fireball mean power outputting from the central engine $\dot{E}$ is a fraction of $L_{\nu\bar{\nu}}$, i.e.,
\beq
\dot{E}=\eta L_{\nu\bar{\nu}},
\eeq
where $\eta$ is the conversion factor \citep[e.g.,][]{Aloy2005,Fan2011,Liu2012b}. The output power can be written as
\beq
\dot{E} \approx \frac {(1+z)(E_{\rm \gamma,iso}+E_{\rm k,iso})\theta_{\rm j}^{2}}{2 T_{90}},
\eeq
where $z$ is the redshift, $E_{\rm \gamma,iso}$ is the isotropic radiated energy in the prompt emission phase, $E_{\rm k,iso}$ is the isotropic kinetic energy of the outflow powering long-lasting afterglow, $T_{\rm 90}$ can roughly be considered as the duration of the activity of the central engine, and $\theta_{\rm j}$ is the opening angle of the ejecta.

Hence, for the cases of $\dot{M}_{\rm ign}<\dot{M}<\dot{M}_{\rm trap}$, we have the mean accretion rate \citep{Fan2011}
\beq
\dot{M} \approx 0.12~[\frac{(1+z)(E_{\rm \gamma,iso,51}+E_{\rm k,iso,51})\theta_{\rm j}^{2}}{\eta T_{90,\rm s}}]^{4/9}~{x_{\rm ms}^{2.1}~(\frac{M_{\rm BH}}{M_\odot}})^{2/3}~M_\odot~\rm s^{-1},
\eeq
where $E_{\rm k,iso,51}=E_{\rm k,iso}/(10^{51}~\rm ergs)$, $E_{\rm \gamma,iso,51}=E_{\rm \gamma,iso}/(10^{51}~\rm ergs)$, and $T_{90, \rm s}=T_{90}/(1~\rm s)$. Furthermore, the disk mass is
\beq
M_{\rm disk} \approx  0.12~[\frac{(E_{\rm \gamma,iso,51}+E_{\rm k,iso,51})\theta_{\rm j}^{2}}{\eta}]^{4/9}~(\frac{T_{90,\rm s}}{{1+z}})^{5/9}~x_{\rm ms}^{2.1}~(\frac{M_{\rm BH}}{M_\odot})^{2/3}~{M_{\odot}}.
\label{eq:M_disk}
\eeq

According to the above equation, we can estimate the disk mass by using the observational data. It should be noted that there exist some uncertainties, especially for the efficiency $\eta$ and the interval of the activity of central engine replaced by $T_{\rm 90}$. There should exist an efficiency from the neutrino annihilation to the initial fireball, then to the jet kinetic energy and radiation, which is mainly related to the energy, components, and state of the fireball \citep[e.g.,][]{Eichler1989,Aloy2005}. \citet{Aloy2005} mentioned that the duration of the GRB event might be longer than the time interval of the activity of central engine if the radial expansion of the fireball is considered. In the fireball model, it is difficult to estimate the duration of such an expansion to the optically thin phase by the observational data unless the blackbody component can be observed. It is conceivable that the consequences of the use of $\eta$ and $T_{\rm 90}$ would change the resulting disk mass to some extent although the exponents of $\eta$ and $T_{\rm 90}$ in Equation (8) are small.

$E_{\rm \gamma,iso}$ can be calculated by the observational data, which is written as
\beq
E_{\rm \gamma,iso}=4\pi D_L^2 F_\gamma /(1+z),
\eeq
where $D_L$ is the luminosity distance, and $F_\gamma$ is the fluence in the 15-150 keV for $\emph{Swift}$ events. Then $D_L$ is defined as
\beq
D_L= \frac{(1+z)c}{H_0} \int_{0}^{z} [\Omega_{\rm M} (1+z')^3 + \Omega_\Lambda]^{-1/2} d z',
\eeq
here we employ a standard $\Lambda$CDM cosmology model with $\Omega_{\rm M} = 0.27$, $\Omega_\Lambda = 0.73$, and $H_0 = 71~\rm km~s^{-1}~Mpc^{-1}$. Moreover, the mean isotropic gamma-ray luminosity is
\beq
L_{\rm \gamma,iso} \approx E_{\rm \gamma,iso}(1+z)/T_{90}.
\eeq

$E_{\rm k,iso}$ and $\theta_{\rm j}$ can be deduced from the modeling of the X-ray afterglow data. We take $E_{\rm k,iso}$ as \citep{Lloyd-Ronning2004,Fan2006,Zhang2007b}
\beq
E_{\rm k,iso} &\approx& 9.2 \times 10^{52} R L_{X,46}^{4/(p+2)}(\frac{1+z}{2})^{-1} \epsilon_{B,-2}^{(2-p)/(p+2)} \epsilon_{\rm e,-1}^{4(1-p)/(p+2)}\nonumber\\ &&\times t_{\rm d}^{(3p-2)/(p+2)} (1+Y)^{4/(p+2)}~{\rm ergs},
\label{eq:E_iso}
\eeq
where $R \sim (t_{11}/T_{90, \rm s})^{17\epsilon_e/16}$ is a factor that accounts for the energy loss during the deceleration following the prompt gamma-ray emission phase \citep[e.g.,][]{Sari1997,Lloyd-Ronning2004}, $\epsilon_{\rm e,0.1}=\epsilon_{\rm e}/0.1$ is the fractions of shock energy given to the electrons, $\epsilon_{\rm B,0.1}=\epsilon_{\rm B}/0.01$ is the fraction of energy in the magnetic field, $t_{11}=t/({\rm 11~hours})$ and $t_{\rm d}=t/(\rm 1~day)$ are the time of observation, $Y$ is Compton parameter, $p$ is the energy distribution index of the shock-accelerated electrons and can be fitted by the observed photon index in the X-ray spectrum \citep[e.g.,][]{Zhang2006b,Gao2013}, and $L_{X,46}={L_X/(10^{46}}~\rm erg~s^{-1})$ is the isotropic X-ray afterglow luminosity. Here we take the X-ray luminosity at 11 hours since the burst triggers, which can be written as
\beq
L_X=4\pi D_L^2 F_X,
\eeq
where $F_X$ is the X-ray flux of the afterglow recorded by satellites.

Furthermore, the relation between the opening angle and the jet  break time is given by \citep[e.g.,][]{Sari1999,Frail2001,Fong2012}
\beq
\theta_{\rm j} \approx 0.076~(\frac{t_{\rm j}}{1~{\rm day}})^{3/8}(\frac{{1+z}}{2})^{-3/8}(\frac{n}{\rm 0.01~cm^{-3}})^{1/8} E_{\rm k,iso,51}^{-1/8},
\label{eq:theta_j}
\eeq
where $t_{\rm j}$ is the jet break time in the X-ray afterglow phase of GRBs, and $n$ is the number density of the burst circumstance.

\section{Results}

The relations among the observational data of the prompt emission and afterglow in SGRBs, disk mass, and BH parameters are established by Equations (8-14). If the reasonable ranges of the BH parameters are given, the limits of the disk masses corresponding to the certain SGRBs can be resolved.

\subsection{Data of SGRBs}

\citet{Berger2014} mainly reviewed the progresses of SGRBs in the theories and observations, including the afterglow and host galaxy observations, the properties of the circumburst environments and their progenitors. There are 70 SGRBs with a substantial fraction of afterglow detections in the eight-year period from January 2005 to January 2013 \citep{Berger2014}, with the addition of GRB 130603B \citep[e.g.,][]{Berger2013,Tanvir2013}, which is associated with a kilonova \citep{Li1998}. As shown in Table 1 of \citet{Berger2014}, there are 27 SGRBs with the authentic X-ray detections and known redshifts discovered by $\emph{Swift}$ satellite except for GRB 050709 by HETE-2.

Moreover, we find 4 SGRBs with the X-ray detections and known redshifts triggered after GRB 130603B, i.e., GRBs 131001A, 140622A, 140903A, and 141212A, whose data are from the UK Swift Science Data Centre \citep{Evans2009}. So totally 31 SGRBs are listed in Table 1. For each SGRBs, we fit the photon index with the data of X-ray spectrum to deduce $p$ \citep[e.g.,][]{Zhang2006b,Gao2013}. Their durations $T_{90}$, redshifts $z$, gamma-ray fluences $F_\gamma$ and X-ray fluxes at 11 hours since trigger $F_X$ (11 hours), and the observed spectral index $\beta$ are displayed. If we take $\epsilon_{\rm e} \sim 0.1$, $\epsilon_{\rm B}\sim 0.01$, $Y \sim 0$, $\eta=0.3$, and fitted $p$ and given $t_{\rm j}$, then $E_{\rm \gamma,iso}$, $E_{\rm k,iso}$, and $\theta_{\rm j}$ can be solved, and we further obtain the ranges of the disk masses.

It is worth noting that the most difficult problem is the estimation of the jet opening angle $\theta_{\rm j}$ because of the faint and restricted observations of SGRB afterglows. So far there are three scenarios, i.e., (1) a few credible detections of jet break, such as in GRBs 051221A \citep{Soderberg2006}, 090426 \citep{Nicuesa2011}, and 130603B \citep{Fong2013}; (2) several meaningful lower limits on jet opening angles, such as in GRBs 050724 \citep{Grupe2006}, 111117A \citep{Margutti2012}, and 120804A \citep{Berger2013}; (3) no break in X-ray lightcurves of some SGRBs. We cite the data of the jet opening angles or their lower limits in the above references for former two cases as shown in Table 1. For the third scenario, we set the lower limit of $\theta_{\rm j} \gtrsim 0.05$ \citep{Fong2012}. For the last 4 SGRBs we collected, the jet break time, $\sim 3.80_{-2.50}^{+0.00}~\rm ks$, is found in GRB 140903, then we can estimate the opening angle by Equation (\ref{eq:theta_j}) with $n = 0.01~\rm cm^{-3}$, and the other three are set by the lower limit as discussed above.

\subsection{Disk masses of SGRBs}

Figure 1 shows the ranges of disk masses $M_{\rm disk}$ of the different SGRBs with the isotropic gamma-ray luminosity $L_{\rm \gamma,iso}$ for varying BH mass $M_{\rm BH}$ from $2.7~M_\odot$ to $10~M_\odot$ and BH spin $a_*$ from 0 to 0.99. Three vertical lines correspond to $M_{\rm disk}= 0.2~M_{\odot},~0.5~M_{\odot}$, and $1~M_{\odot}$, respectively. It is seen that $M_{\rm disk}$ has a wide distribution, from about $6 \times 10^{-4}~M_{\odot}$ to about $7.6~M_{\odot}$, as well as shown in Table 1. The accretion rates corresponding to the minimal disk masses are checked, which are in the suitable ranges, i.e., $\dot {M}_{\rm ign} < \dot{M} < \dot{M}_{\rm trap}$. There is no statistical correlation between the disk mass and gamma-ray isotropic luminosity, because the energy coming from the accretion powers all radiative processes of GRBs, mainly in gamma-ray and X-ray bands. The energies of the X-ray afterglows are frequently larger than those of prompt emission as displayed in Table 1. Obviously, the disk mass is primarily calculated by the output energy of GRBs, opening angle of the jet, and BH characteristics as shown in Equation (8). There exists a difference of several orders of magnitude between the minimal and maximal disk mass, which means that the BH characteristics are the major factors on the disk mass.

As shown in the figure, the maximal disk mass of GRBs 050724, 051221A, 070714B, 070809, 090426, 111117A, 120804A, and 131001A are larger than $1~M_\odot$, and that of most other SGRBs in our sample are larger than $0.2~M_\odot$, which indicates that the extreme BH spin parameters and small BH mass are required. In binary NS merger events, the BH mass is naturally less than the total mass of the binary, i.e., $\sim 4~M_\odot$, which is described about $\sim 2.7~M_\odot$ in further simulations. In BH-NS binaries, the BH is origin from its progenitor star, and its mass should also be a stellar-mass order, e.g., $\sim 10~M_\odot$. Moreover, the BH spin parameters are also related to their progenitors. In some discussions \citep[e.g.,][]{Lee2000b,Ruffert2001}, the rapidly rotating BHs ($a_* \geq 0.5$) are inclined to exist in the SGRB centers in contrast to the BHs in the LGRB centers.

In order to embody the effects of the BH characteristics on the disk mass, Figure 2 displays the distributions of the disk masses $M_{\rm disk}$ for the different typical BH masses and spins, which are set to ($M_{\rm BH}/{M_\odot}$, $a_*$) = (3, 0.5), (3, 0.9), (10, 0.5), and (10, 0.9), corresponding to (a-d), respectively. It is easy to find that the disk mass of most SGRBs are safe below $0.2-0.4~M_\odot$, and sporadic cases are beyond the limits, especially in the case of Figure 2(c). Additionally, by comparing these four cases, we notice that the spin parameters are more effective than the BH masses on the values of the disk masses. Even for the case of Figure 2(b), there still exists one SGRB, whose disk mass is larger than $0.45~M_\odot$. Those massive disks, $\geq 1~M_\odot$, may exist in the centers of collapsars, which are considered as the origin of LGRBs. Actually, \citet{Lazzati2010} proposed that the off-axis jets from collapsars could power SGRBs.

Three factors remind us that the results of the disk masses are the lower limits at the most, which are as follows. (1) We have to calculate the disk mass using the lower limit of $\theta_{\rm j}$ in most SGRBs as shown in Table 1. It is easily conceivable that the real requirements of the disk masses are much larger than the the present results if the precise value of $\theta_{\rm j}$ is considered. (2) Some powerful SGRBs with unknown redshift, such as GRBs 060121, 060313, and 111121A, shown in Table 1 of \citet{Berger2014}, may require more massive disk than SGRBs in our sample if they are also origin from the BH hyperaccretion systems. (3) The powerful X-ray flares have been extensively observed in the afterglow phase of GRBs, which are considered to originate from the re-ignition of the central engine \citep[e.g.,][]{Liu2008,Luo2013,Hou2014}. This means that the remanent matters from the massive disk are needed to maintain the explosion of X-ray flares. However, we use $T_{\rm 90}$ to replace the duration of the activity of the central engine, which may generally enlarge the disk mass in the calculations. Although these influences and some uncertainties may exist, we consider that our results can still reflect the deficiency of the neutrino annihilation process to power SGRBs.

\section{Conclusions and discussion}

The progenitor of SGRBs is considered to be a compact binaries merger event. After merger, a stellar-mass BH surrounded by an NDAF will be formed in the central SGRB and neutrino-antineutrino annihilation above the disk may power SGRBs. The total mass of two compact stars limits the mass of the system consisting of a BH and an NDAF. In this paper, we focus on a question, i.e., can annihilations of neutrinos from NDAFs owning such masses power all the observed SGRBs? The calculations show that the disk mass of a certain SGRB mainly depend on its output energy, jet opening angle, and central BH characteristics. Even for the extreme BH parameters, there still exist some SGRBs requiring the massive disks, which approach or exceed the limits in simulations.

Besides magnetar model, for BH hyperaccretion system, we suggest that there may exist an alternative magnetic origin of SGRBs, i.e., BZ process \citep[e.g.,][]{Popham1999,Lee2000a,Lee2000b,DiMatteo2002,Kawanaka2013} or episodic magnetic reconnection \citep{Yuan2012}, to replace neutrino annihilation. \citet{Kawanaka2013} presented that the luminosity powered by Poynting-dominated jet is more qualified for the requirement of GRBs than neutrino pair annihilation. \citet{Yuan2012} investigated that the closed magnetic field lines continuously emerge out of the accretion flow. Since the shear and turbulent motion of the accretion flow, the line may form the flux rope. When a threshold is reached, the system loses its equilibrium and the flux rope is thrust outward, then an episodic jet occurs. This mechanism can also power enormous energy to trigger GRBs. In addition, if these magnetic origins really exist in the center of GRBs, the polarization effect should be observed in the prompt emission or afterglow of GRBs. Actually, the linear polarization in the afterglow of LGRB GRB 120308 has been detected \citep{Mundell2013}, which indicates that large-scale magnetic fields may be dominant in the GRB jets. But now we do not know whether the same situation exists in SGRBs.

Otherwise, there are some mechanisms, such as magnetic coupling from the BH horizon to the inner region of the disk \citep{Li2000}, can effectively transfer the angular momentum and rotational energy of the BH to heat the inner region of the disk, then larger numbers of neutrinos radiate from the disk to produce the primordial fireball \citep[e.g.,][]{Lei2009,Luo2013}. Besides, the vertical advection (or convection) is considered to widely exist in the slim disks and NDAFs \citep{Jiang2014,Liu2015}, which is another possible mechanism to increase the neutrino emission rate. The scenario for NDAFs follows, the vertical advection (or convection) caused by magnetic buoyancy can much effectively transport energy to the disk surface, and also suppress the radial advection, thus the neutrino luminosity and annihilation luminosity are dramatically increased. This mechanism is conducive to achieve the energy requirement of GRBs.

\begin{acknowledgements}
We thank Yi-Zhong Fan, Fu-Wen Zhang and Xiao-Hong Zhao for beneficial discussion and the anonymous referee for very useful suggestions and comments. We acknowledge the use of the public data from the \emph{Swift} archives. Our work also made use of data supplied by the UK Swift Science Data Centre at the University of Leicester. This work was supported by the National Basic Research Program of China (973 Program) under grant 2014CB845800, the National Natural Science Foundation of China under grants 11222328, 11233006, 11333004, 11373002, 11473022, U1231101, and U1331101, and the CAS Open Research Program of Key Laboratory for the Structure and Evolution of Celestial Objects under grants OP201305 and OP201403.
\end{acknowledgements}

\clearpage

\begin{table}
\tiny
\caption{Data of SGRBs}
\begin{center}
\begin{tabular}{lllcccccclc}
\hline\hline
& GRB & $T_{90}$ & $z$ & $F_{\rm \gamma }          $ & $F_{\rm X}(11~\rm hours)            $  & Photon index & $E_{\rm \gamma,iso}$ & $E_{\rm k,iso}$ $^a$ & $~~~~\theta_{\rm j}$ & $M_{\rm disk}$$^b$\\
&     & (s)      &     & $(10^{-7}~\rm erg~cm^{-2})$ & $(10^{-14}$ $\rm erg~cm^{-2}~s^{-1})$  &              & $(10^{51}~\rm ergs)$ & $(10^{51}~\rm ergs)$ & ~~(rad)              & $(M_{\odot})$ \\
\hline
&050509B & 0.04    &  0.225     &  0.23   &  $<$ 1.95  & $1.6_{-0.4}^{+0.5}$    & 0.0027  & 0.055  & $\gtrsim 0.05$       &  0.0006-0.028\\
&050709  & 0.07    &  0.161     &  4.0    &  1.92      & $\sim 2$               & 0.023   & 0.016  & $\gtrsim 0.26$ (1)   &  0.003-0.14\\
&050724  & 3       &  0.257     &  6.3    &  9.55      & $1.68_{-0.13}^{+0.15}$ & 0.1     & 0.27   & $\gtrsim 0.35$ (2)   &  0.084-3.93\\
&051210	 & 1.3	   &  1.3       &  0.83   &  $<$ 2.7   & $2.78_{-0.41}^{+0.48}$ & 0.36	  & 2.38   & $\gtrsim 0.05$       &  0.016-0.76\\
&051221A & 1.4     &  0.5465    &  12     &  108       & $2.09_{-0.09}^{+0.10}$ & 0.92    & 12.6   & $\sim 0.12$ (3)      &  0.093-4.37\\
&060502B & 0.09    &  0.287     &  0.4    &  $<$ 1.47  & $2.15_{-0.58}^{+1.07}$ & 0.012   & 0.12   & $\gtrsim 0.05$       &  0.0013-0.062\\
&060801  & 0.5     &  1.130     &  0.81   &  $<$ 0.98  & $2.01_{-0.26}^{+0.23}$ & 0.27    & 0.71   & $\gtrsim 0.05$       &  0.0063-0.30\\
&061006  & 0.4     &  0.438     &  14     &  22.7      & $1.86_{-0.24}^{+0.30}$ & 0.67    & 3.14   & $\gtrsim 0.05$       &  0.013-0.59\\
&061201  & 0.8     &  0.111     &  3.3    &  19.2      & $1.54_{-0.17}^{+0.17}$ & 0.01    & 0.07   & $\sim 0.017$ (4)     &  0.0014-0.067\\
&061210	 & 0.2	   &  0.409     &  3.0    &  13.6      & $2.60_{-0.71}^{+1.92}$ & 0.12	  & 0.86   & $\gtrsim 0.05$       &  0.0048-0.22\\
&070429B & 0.5     &  0.902     &  0.63   &  11.3      & $2.69_{-0.56}^{+1.18}$ & 0.13    & 4.51   & $\gtrsim 0.05$       &  0.013-0.63\\
&070714B & 2.0     &  0.923     &  7.2    &  6.30      & $1.96_{-0.15}^{+0.12}$ & 1.61    & 2.32   & $\gtrsim 0.05$       &  0.027-1.25\\
&070724A & 0.4     &  0.457     &  0.30   &  12.8      & $1.46_{-0.25}^{+0.36}$ & 0.016   & 0.99   & $\gtrsim 0.05$       &  0.007-0.33\\
&070729  & 0.9     &  0.8       &  1.0    &  $<$ 4.71  & $1.5_{-0.3}^{+0.6}$    & 0.17    & 1.32   & $\gtrsim 0.05$       &  0.012-0.54\\
&070809	 & 1.3     &  0.473     &  1.0    &  53.0	   & $1.39_{-0.12}^{+0.14}$ & 0.056   & 3.91   & $\gtrsim 0.05$       &  0.024-1.15\\
&071227  & 1.8     &  0.381     &  2.2    &  3.20      & $2.19_{-0.35}^{+0.41}$ & 0.08    & 0.25   & $\gtrsim 0.05$       &  0.01-0.47\\
&080905A & 1.0	   &  0.122     &  1.4    &  $<$ 6.7   & $1.54_{-0.14}^{+0.22}$ & 0.005   & 0.024  & $\gtrsim 0.05$       &  0.0027-0.13\\
&090426  & 1.2     &  2.609     &  1.8    &  26.3      & $2.03_{-0.15}^{+0.16}$ & 2.84    & 135    & $\sim 0.07$ (5)      &  0.093-4.35\\
&090510  & 0.3     &  0.903     &  3.4    &  5.04      & $1.70_{-0.12}^{+0.12}$ & 0.73    & 3.07   & $\sim 0.017$ (4)     &  0.0035-0.17\\
&090515	 & 0.04	   &  0.403     &  0.21   &  $<$ 8.43  & $2.73_{-0.77}^{+1.20}$ & 0.008   & 0.62   & $\gtrsim 0.05$       &  0.0016-0.075\\
&100117A & 0.3	   &  0.915     &  0.93   &  $<$ 2.50  & $2.74_{-0.31}^{+0.36}$ & 0.20    & 1.10   & $\gtrsim 0.05$       &  0.0057-0.27\\
&100206A & 0.1     &  0.408     &  1.4    &  $<$ 1.07  & $2.0_{-0.7}^{+0.8}$    & 0.058   & 0.073  & $\gtrsim 0.05$       &  0.0013-0.062\\
&100625A & 0.3	   &  0.453     &  2.3    &  0.395     & $2.3_{-0.3}^{+0.5}$    & 0.12    & 0.093  & $\gtrsim 0.05$       &  0.003-0.14\\
&101219A & 0.6	   &  0.718     &  4.6    &  2.00      & $1.44_{-0.25}^{+0.27}$ & 0.62    & 0.45   & $\gtrsim 0.05$       &  0.008-0.38\\
&111117A & 0.5     &  1.3       &  1.4    &  3.21      & $2.10_{-0.32}^{+0.39}$ & 0.62    & 3.77   & 0.105 (6)            &  0.023-1.06\\
&120804A & 0.81	   &  1.3       &  8.8    &  58.6      & $2.10_{-0.14}^{+0.22}$ & 3.88	  & 56.9   & $\gtrsim 0.19$ (7)   &  0.16-7.59\\
&130603B & 0.18    &  0.356     &  6.3    &  60.0	   & $2.00_{-0.13}^{+0.14}$ & 0.20    & 2.80   & $\sim 0.07$ (8)      &  0.01-0.48\\
&131001A & 1.54    &  0.717     &  2.8    &  14.7      & $1.91_{-0.18}^{+0.18}$ & 0.37    & 5.41   & $\gtrsim 0.05$       &  0.029-1.37\\
&140622A & 0.13    &  0.959     &  0.27   &  17.0      & $1.55_{-0.28}^{+0.67}$ & 0.065   & 9.77   & $\gtrsim 0.05$       &  0.0087-0.41\\
&140903A & 0.30    &  0.351     &  1.4    &  124.7     & $1.59_{-0.20}^{+0.22}$ & 0.043   & 6.15   & 0.023 (9)            &  0.0067-0.31\\
&141212A & 0.30    &  0.596     &  0.72   &  2.50      & $2.0_{-0.5}^{+0.8}$    & 0.066   & 0.38   & $\gtrsim 0.05$       &  0.0039-0.18\\
\hline
\hline
\end{tabular}
\end{center}
\begin{minipage}{16cm}
\emph{Notes}:\\
$^a$ The parameters are calculated by Equation (\ref{eq:E_iso}) with $\epsilon_{\rm e} \sim 0.1$, $\epsilon_{\rm B}\sim 0.01$, and $Y \sim 0$. \\
$^b$ The ranges of $M_{\rm disk}$ are estimated by Equation (\ref{eq:M_disk}) with $\eta=0.3$, varying $M_{\rm BH}$ from $2.7~M_\odot$ to $10~M_\odot$, and $a_*$ from 0 to 0.99.\\
\emph{References}: \\
(1) \citealt{Berger2014}; (2) \citealt{Grupe2006}; (3) \citealt{Soderberg2006}; (4) \citealt{DePasquale2010,Nicuesa2012}; (5) \citealt{Nicuesa2011}; (6) \citealt{Margutti2012}; (7) \citealt{Berger2013}; (8) \citealt{Fong2013}; (9) The opening angle of GRB 140903A is determined by Equation (\ref{eq:theta_j}) with the data from the UK Swift Science Data Centre \citep{Evans2009} and $n \sim 0.01~{\rm cm^{-3}}$.\\
\end{minipage}
\label{tab:sample}
\end{table}

\clearpage

\begin{figure*}
\centering
\includegraphics[angle=0,scale=0.6]{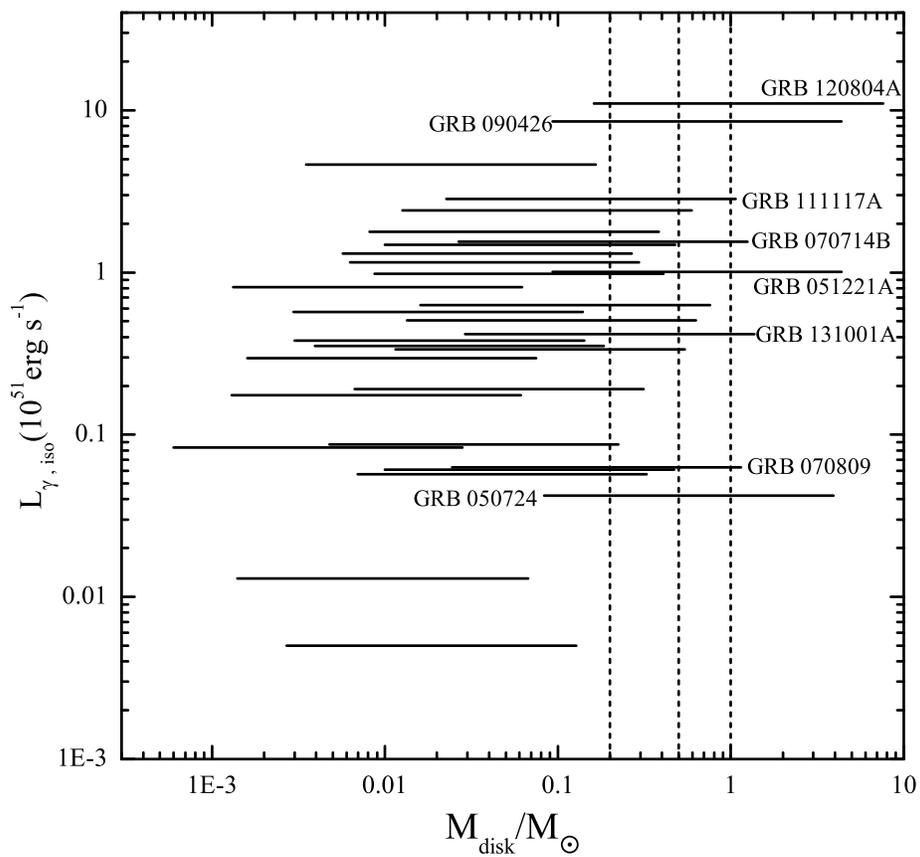}
\caption{Ranges of disk masses $M_{\rm disk}$ of different SGRBs with isotropic gamma-ray luminosity $L_{\rm \gamma,iso}$ for varying BH mass $M_{\rm BH}$ from $2.7~M_\odot$ to $10~M_\odot$ and BH spin $a_*$ from 0 to 0.99. Three vertical lines correspond to $M_{\rm disk}= 0.2~M_{\odot},~0.5~M_{\odot}$, and $1~M_{\odot}$, respectively.}
\label{fig1}
\end{figure*}

\clearpage

\begin{figure*}
\centering
\includegraphics[angle=0,scale=0.7]{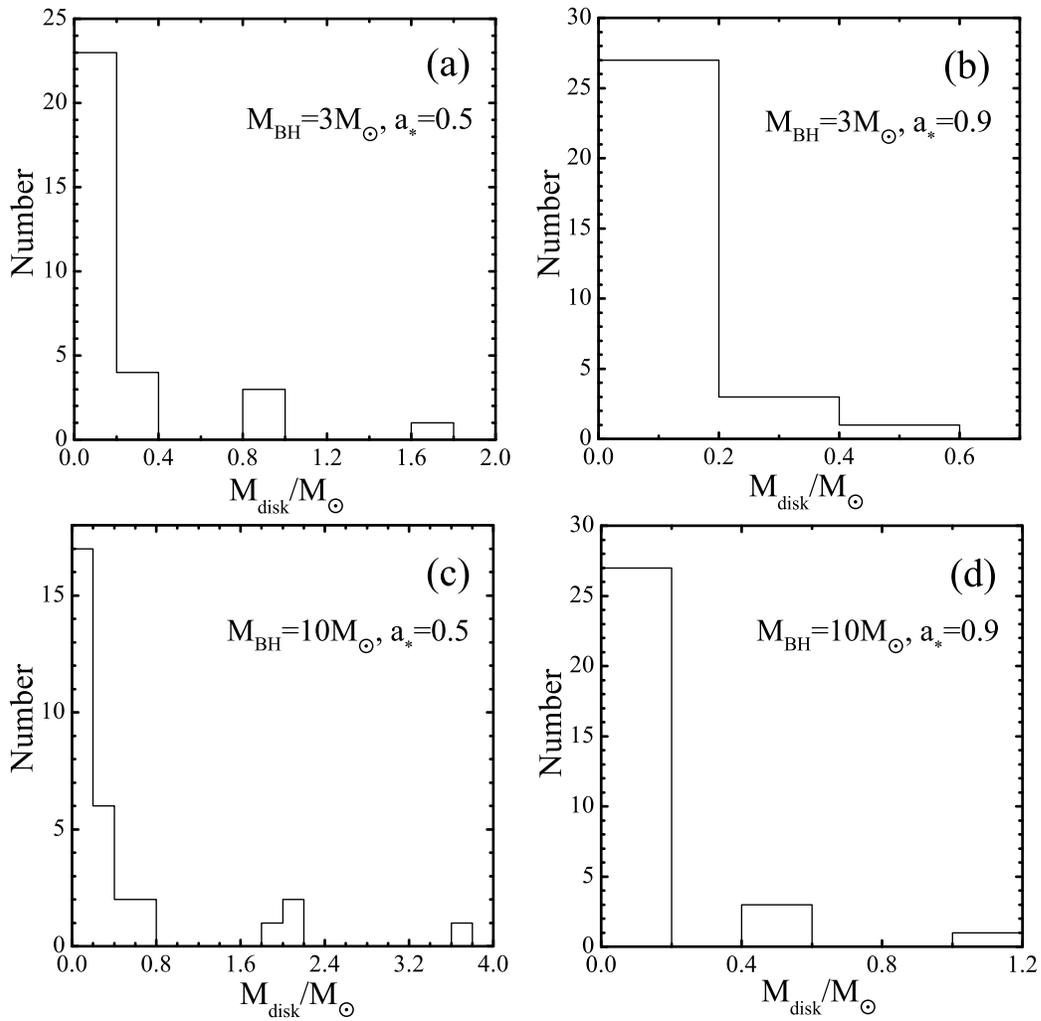}
\caption{Distributions of the disk masses $M_{\rm disk}$ for different typical BH masses and spins.}
\label{fig2}
\end{figure*}

\clearpage

\end{document}